\documentclass[pre,twocolumn]{revtex4}
\usepackage{graphicx}
\usepackage{amsmath}
\usepackage{amssymb}
\usepackage{color}
\usepackage[normalem]{ulem}
\usepackage{epsfig}

\begin{document}
\bibliographystyle{apsrev4-1}

\title{The effect of environmental stochasticity on species richness in neutral communities}

\author{Matan Danino$^1$, Nadav M. Shnerb$^1$, Sandro Azaele$^2$, William E. Kunin$^3$  and David A. Kessler$^1$ }

\affiliation{$^1$ Department of Physics, Bar-Ilan University, Ramat-Gan
IL52900, Israel \\ $^2$ Department of Applied Mathematics, School of Mathematics, University of Leeds, Leeds, UK \\ $^3$ School of Biology, University of Leeds, Leeds, UK.}

\begin{abstract}
Environmental stochasticity is known to be a destabilizing factor, increasing abundance fluctuations and extinction rates of populations. However, the stability of a community may benefit from the differential response of species to environmental variations due to the storage effect. This paper provides a systematic and comprehensive discussion of these two contradicting tendencies, using the metacommunity version of the recently proposed time-average neutral model of biodiversity which incorporates environmental stochasticity and demographic noise and allows for extinction and speciation. We show that the incorporation of demographic noise into the model is essential to its applicability, yielding realistic behavior of the system when fitness variations are relatively weak. The dependence of species richness on the strength of environmental stochasticity changes sign when the correlation time of the environmental variations increases. This transition marks the point at which the storage effect no longer succeeds in stabilizing the community.
\end{abstract}
\maketitle

\section{Introduction}

One of the biggest puzzles in community ecology is the persistence of high-diversity assemblages. The competitive exclusion principle \cite{gause2003struggle,hardin1960competitive}
 predicts that the number of species coexisting in a local community should be few than or equal to the number of limiting resources, in apparent contrast with the dozens and hundreds of locally coexisting species of  freshwater plankton \cite{hutchinson1961paradox,stomp2011large}, trees in tropical forests
\cite{ter2013hyperdominance} and coral reef species
\cite{connolly2014commonness}.  This problem has received considerable attention in recent decades, with many mechanisms  suggested to circumvent the mathematical constraints embodied in the exclusion principle and many works that try to provide empirical support to these theories \cite{chesson2000mechanisms}.

Within this framework, neutral theories, and in particular the neutral theory of biodiversity (NTB) suggested by Hubbell~\cite{Hubbell2001unifiedNeutral,maritan1,TREE2011}, play an important role. Under neutral dynamics all individuals are considered as having the same fitness, and abundance variations are the result of demographic noise alone.
The number of individuals belonging to each species varies randomly within the limit imposed by the overall size of the community, with most populations eventually drifting to extinction. However, the neutral turnover rate is very slow, and diversity is maintained due to the introduction of new species into the system, either via speciation (in the metacommunity) or by migration (in a local community).

 The slow turnover dynamics in the neutral model is not only an advantage, it is also a disadvantage, and has triggered one of the main lines of criticism directed at the neutral model. It turns out that pure ecological drift is far too slow to account for both the observed short-term fluctuations and the long term dynamics~\cite{ricklefs2006unified,nee2005neutral,leigh2007neutral,kalyuzhny2014temporal,kalyuzhny2014niche,chisholm2014temporal}. For example, the abundance of the most common species in the Barro-Colorado Island Smithsonian 50 ha plot has decreased from 40000 to 30000 individuals ($>1cm \ \rm{dbh}$) during about half of a generation, while under pure demographic noise one expects variations of order $\sqrt{N} \sim 200$ within a whole generation. The abundance of the most common species in the Amazon basin is about $10^9$ individuals~\cite{ter2013hyperdominance}, which sets the time (in generations) needed for their development from a single tree under neutral dynamics, but since a generation time is about 50 years, this timescale is longer than the lifetime of the universe.

Motivated by these difficulties, recent works~\cite{kalyuzhny2015neutral, kessler2014neutral} have pointed towards a generalized neutral theory that will include  both demographic and \emph{environmental} stochasticity. Basically, this new model accepts the equivalence principle, but assumes that the fitness of all species is equal \emph{when averaged over time}, while at any instant some species have higher fitness than the others due to temporal variations in parameters such as temperature, precipitation etc. The ability of this  time-averaged neutral theory of biodiversity (TNTB) to explain different empirical patterns, including species abundance distributions, temporal fluctuations statistics  and the growth in system dissimilarity over time, was demonstrated in \cite{kalyuzhny2015neutral}.

However, by introducing a species-specific response to environmental variations, the TNTB finds itself entering the domain of another celebrated mechanism that was suggested to explain species coexistence, the storage effect introduced by Chesson in the 1980s. In particular, Chesson and Warner~\cite{chesson1981environmental} considered the ``lottery game" in which the fitness of each species, as reflected by the chance of its offspring successfully occupying a vacancy in the community, fluctuates in time.  This differential response of species, when superimposed on buffered population growth and covariance between environment relative probability and competition~\cite{chesson1994multispecies} was shown to stabilize the system. Chesson and Warner showed how rare species, when compared with common species, have fewer per-capita losses when their fitness is low and more gains when their fitness is high. Accordingly, the population of rare species increases (their average growth rate is positive just because their relative abundance is low) and the system supports a stable equilibrium.

 Hubbell's NTB, which takes into account demographic noise and speciation but with no environmental noise, provides us with one set of predictions for the patterns characterizing a community, such as species abundance distribution and species richness. The Chesson-Warner lottery game, taking into account only environmental stochasticity (without demographic noise or speciation) suggests another set. What happens under the general model of TNTB, where \emph{all} these elements play a role?  What patterns does it predict, and how do they depend on the strength of the storage effect? In \cite{kalyuzhny2015neutral} the TNTB was presented in the context of a mainland-island model and simulated island dynamics were compared with data from the Barro-Colorado Island (BCI) plot. Here we aim at understanding the metacommunity dynamics of the TNTB and to explain its relationships with both NTB  and the lottery game.

To do that, we first revisit the storage effect, using the original Chesson-Warner model. In section II we consider the storage effect for two species, emphasizing the transition it shows from a balanced system, where the abundance of both species fluctuates around one half of the community, and an imbalanced state, with one rare and one frequent species. A deeper analysis of the equilibrium distribution poses a conceptual problem, namely that the result is independent of the amplitude of the environmental variations. This problem is discussed in section III, and indicates the need to incorporate demographic stochasticity into the model. Before doing that, in section IV we consider the original lottery game  for communities with many species and discuss its applicability to empirical systems. Finally in section V the TNTB model, in which environmental variations, demographic stochasticity and speciation affect the community, is analyzed. Conclusions are presented in the last section.

\section{A lottery game for two species}

In this section we study the simplest case, the storage effect in a community with two species playing the lottery game. Since we are ultimately interested in the TNTB, we assume that the fitness of both species is equal when averaged over time (species are equivalent). Note that the scope of the storage effect is wider, and it may stabilize a community even when the average fitness is different; we will return to this point in the discussion section.

To provide an intuitive numerical example, let us consider an extremely simple game.  Imagine a forest with 100 trees, $N_A$ of species A and $N_B = 100-N_A$ of species B. For simplicity we assume that there is no spatial structure, seeds and seedlings of both species are all around the forest, with relative frequencies that reflect the relative abundance of adult trees. During every year $20\%$ of the trees are selected at random, independent of their species affiliation, to die (so that the generation time is five years). The gaps that remain after the trees' death are filled by seedlings, where the chance of each seedling to capture the vacancy depends of its relative fitness, with the fitness  varying in time.  To have equivalent species  the temporal fitness is taken to be an iid variable, so the chance of a particular species to be the fitter of the two in a certain year is $1/2$.  Under an extreme, ``winner takes all" scenario, the fittest species of a given year captures all the 20 empty slots.

Now let us follow the dynamics. Consider the case where, at the beginning of a certain year, $N_A = 20$ and $N_B = 80$. After the death step, $N_A = 16$ and $N_B = 64$ (this is an average, since trees are picked to die at random, but for our purpose it is sufficient to trace the average). Now there are two options: if the winner of this year is species $A$, the year ends with $N_A = 36, \ N_B= 64$, while if the fittest species is $B$, the outcome will be $N_A = 16, \ N_B = 84$. One can easily see that the gain of  $A$ when it wins, $16$, is higher than the potential gain of $B$, which increases its population only by four individuals when it wins. By the same token the losses of $A$ when it is the inferior species are smaller then the losses of $B$ in the parallel situation.

While this example is misleading in several respects (in particular the unrealistic winner takes all assumption strongly affects  the results), it still provides the basic intuition: although the average fitness of both species is the same, environmental variations provide benefit to the rarer one, as the opportunities for the rare species (when it wins) are greater than those of a common species and its risks (when it loses) are less. Accordingly, an effective stabilizing force acts against any deviation from the $50-50$ partition.

Having established this intuition, let us turn to the original two-species model as presented in~\cite{chesson1981environmental}. In this model there is no demographic noise, so the absolute number of individuals has no importance. Accordingly, the variables are species relative fractions. For two species, these are $x_1$ and $1-x_1$.

The model has a two step dynamic. During the death step a fraction $\delta$ of the trees are removed, so the loss of species number 1, for example, is $\delta \cdot x_1$. The gaps are filled by seedlings. The number  of seedlings for a species is proportional to its abundance, and the chance of a single seedling to capture the empty slot is determined by its  species' fitness. Accordingly the abundance of the two species after these steps, death and recruitment, is given by,
\begin{eqnarray} \label{eq1}
x_1^{t+1} = x_1^{t} (1-\delta) + \delta \frac{f_1^t x_1^t}{f_1^t x_1^t + f_2^2 x_2^t} \nonumber \\
x_2^{t+1} = x_2^{t} (1-\delta) + \delta \frac{f_2^t x_2^t}{f_1^t x_1^t + f_2^t x_2^t}
\end{eqnarray}
where $f_i^t>0$ is the fitness of the $i$-th species  during the $t$-th step. For two species system one can replace $x_2$ by $1-x_1$ to get a single recursion equation, $$x_1^{t+1} = x_1^{t} (1-\delta) + \frac{\delta f_1^t x_1^t}{f_1^t x_1^t + f_2^t (1-x_1^t)}.$$

When the fitness is fixed in time, the fittest species will win the game and the abundance of the inferior species decreases monotonically towards zero. Chesson discovered that, when the fitness fluctuates in time, it stabilizes the populations. As mentioned above, to extend the neutral theory we require the long-term average of $f_1$ and $f_2$ to be equal.

Two parameters are needed to characterize environmental stochasticity: its strength and its duration (correlation time).

\begin{enumerate}
  \item The \emph{strength} of the environmental stochasticity, $\sigma_E^2$ manifests itself in the spread of the fitness parameters $f_i$. Without loss of generality one may take
      \begin{equation} \label{eq0}
      f_i = e^{\gamma_i},
       \end{equation}
       where the parameter $\gamma_i$ is an iid variable picked from a distribution (say, a Gaussian or a uniform distribution) with zero mean and  variance $\sigma_E^2$. If  $\sigma_E^2 = 0$ then $f_i \equiv 1$, all species have the same fitness and the dynamics stops, $x^{t+1}_i = x^t_i$. The larger $\sigma_E^2$ is, the stronger is the fitness variability.

  \item $\delta$ is the \emph{correlation time } of the environmental noise, measured in units of generations. Our analysis of Eqs. (\ref{eq1}) is based on the assumption that the fitness $f_i$ is picked at random every elementary timestep, i.e., between $t$ and $t+1$ and so on. Within this period a fraction $\delta$ of the individuals  die. To give a concrete example, in \cite{kalyuzhny2015neutral} the correlation time of the environmental stochasticity in the Barro-Colorado Island plot was found to be about 10 years, while the generation time is about 50y. To model this dynamics using  Eqs. (\ref{eq1}) one may take  $\delta = 1/5$, meaning  that the replacement of $1/5$ of the trees takes place under (more or less) the same fitness regime.
\end{enumerate}

In \cite{hatfield1989diffusion}, Hatfield and Chesson showed how to map the discrete time equations (\ref{eq1}) to a Fokker-Planck equation for $P(x_1)$, the probability that the relative abundance of species number 1 is $x_1$,
\begin{widetext}\begin{equation} \label{eq2}
 \frac{\partial P(x_1,t)}{\partial t} = \delta \sigma_E^2 \left\{  \frac{\partial }{\partial x_1} \left[x_1(1-x_1)(x_1-1/2)P(x_1,t)
\right] + \delta \frac{\partial^2 }{\partial x_1^2} \left[x_1^2(1-x_1)^2 P(x_1,t)\right] \right\}.
\end{equation}\end{widetext}
The steady-state solution can be seen to be,
\begin{equation} \label{eq3}
P_{eq}(x_1) = C[x_1(1-x_1)]^{\frac{1}{\delta}-2},
\end{equation}
where $C = \Gamma(2/\delta - 2)/\Gamma^2(1/\delta - 1)$.

\begin{figure}
\vspace{-2cm}
\includegraphics[width=7cm]{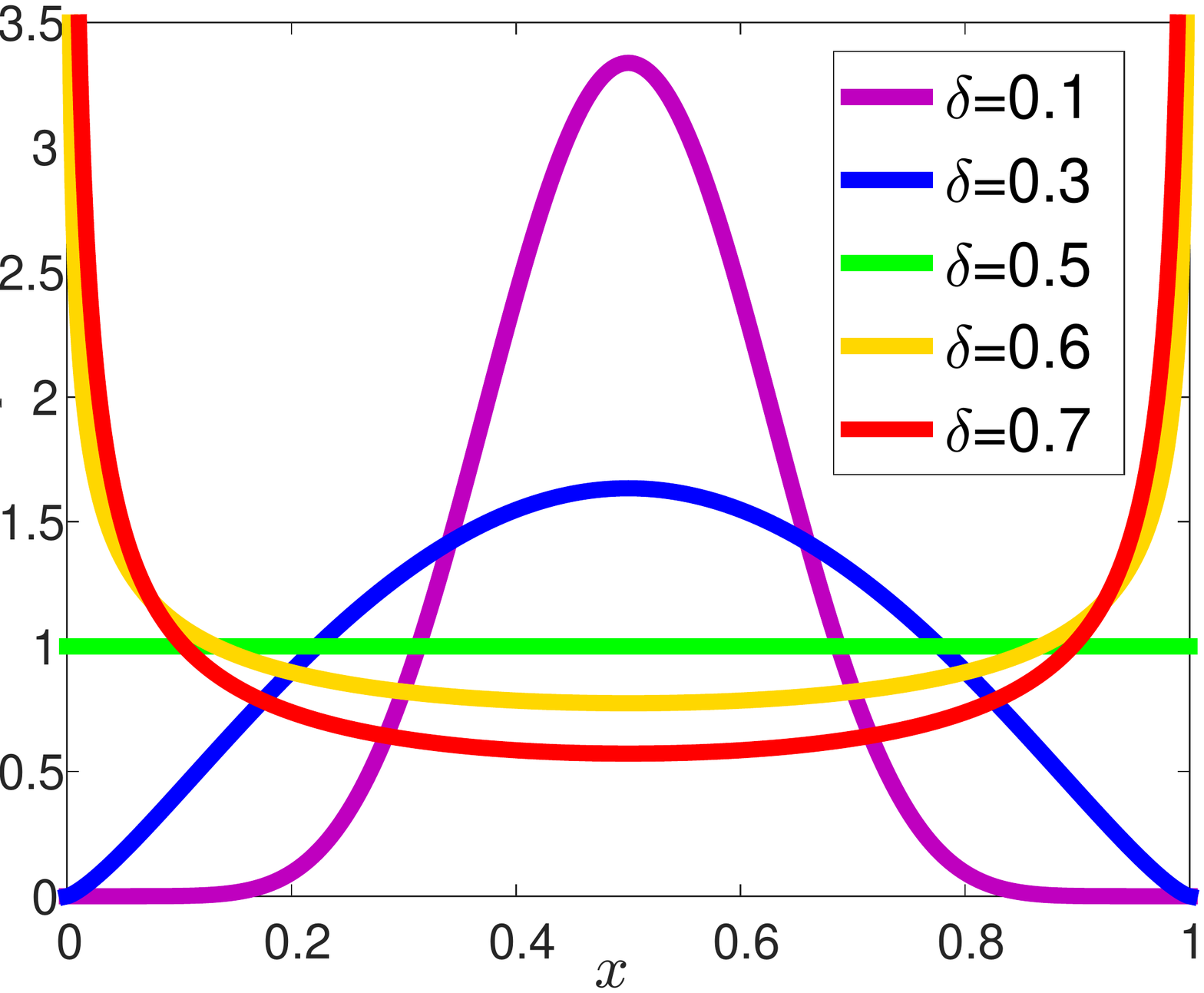}
\vspace{-2cm}
\caption{The equilibrium probability distribution function, $P_{eq}(x)$, given by Eq. (\ref{eq3}), is plotted against $x$ for different values of $\delta$ (see legend). For $\delta<1/2$ the distribution peaks around the symmetric point $0.5$, and the peak becomes sharper when $\delta$ decreases (still the decay is slower than exponential). At $\delta_c =1/2$ the distribution is flat, and for smaller values of $\delta$ it develops two peaks close to the extinction and the fixation points and a valley in the middle. The distribution is normalizable as long as $\delta<1$. However, had the dynamics of Eq. (\ref{eq1}) allowed an absorbing state (e.g., if one consider any fraction $x$ smaller than $x_{min}$ as the state with no individual) the chance of extinction in case of $\delta \ll 1/2$ would have been much smaller than the chance if $\delta > 1/2$. As discussed in the main text, the pdfs shown here are independent of $\sigma_E^2$.  }  \label{fig1}
\end{figure}

This result emphasizes two general features of the storage effect.
First, the right hand side of the Fokker-Planck equation (\ref{eq2}) has two terms. The first is the ``drift" term, describing the dynamics of the average value of $x_1$, which drives $x_1$ towards $1/2$. The ``diffusive" term,  involving the second derivative with respect to $x_1$, has the coefficient $[x_1(1-x_1)]^2$, meaning that the random wandering of the system is strongest when $x_1=1/2$ and approaches zero at the edges, $x_1 = 0$ and $x_1=1$. As discussed in \cite{ohkubo2008transition}, the resulting  $P_{eq}(x_1)$ reflects the balance between these two opposing forces:  the diffusive aspects of the dynamics acts to trap the system close to the edges where the ``diffusion constant" associated with abundance fluctuations vanishes (the model has no demographic noise, and the step size is proportional to $x$ even for vanishingly small values), while the drift term pushes $x_1$ to the stable fixed point in the middle.

The net result is determined by the ratio between these terms, i.e., by $\delta$, as illustrated in Figure \ref{fig1}: for $\delta < 1/2$ the deterministic term wins, leading to a distribution with a single maximum at $1/2$, meaning that at any instant of time the community is likely to be well  balanced, with both species represented by roughly the same number of individuals. For $\delta > 1/2$, on the other hand, $P_{eq}$ is convex, with a lot of support close to the edges at zero and one. In this case the community is unbalanced, (almost) any snapshot picture of the community reveals strong dominance of one species, although the equivalence ensures that the time average fraction of each species is around $1/2$.

The SAD peak for $\delta<1/2$ resembles the Gaussian or exponential peak one finds when the system supports a deterministic stable equilibrium (an attractive fixed point of a nonlinear system), but this similarity is slightly misleading. The decay of $P_{eq}$ towards the edges is described by a power-law, not by an exponential or a Gaussian. This happens since the fixed point at $1/2$ is \emph{noise induced} in the first place.

The second key feature of Eq. (\ref{eq3}) is that it has been derived from Eq. (\ref{eq1}) by expanding it to the leading order in the fitness differences, so the emerging $P_{eq}$ has to be \emph{independent} of $\sigma_E^2$. This approximation becomes better and better as $\sigma_E^2$ decreases;  accordingly, the storage effect appears to  stabilize the two-species community even for vanishingly small values of $\sigma_E^2$. The amplitude of fitness variations only sets the time scale, such that the time needed for the system to reach the equilibrium distribution scales like $1/\sigma_E^2$ and diverges when the environmental noise vanishes, but $P_{eq}$ itself stays the same. In the next section we discuss the conceptual difficulties associated with this outcome.

For the sake of completeness we note that the storage effect acts to destabilize the system when the environmental stochasticity acts on mortality rather than on fecundity. One may realize that easily by repeating the ``loser loses all" version of the numerical example above, where the low fitness species suffers all the mortalities but the recruitment depends on abundance and is independent of fitness. Under such a scenario the rarer species has more to lose than the commoner one, as a the fixed number of deaths would represent a larger share of its population \cite{hatfield1989diffusion}.

\section{Storage effect and demographic stochasticity: a conceptual discussion}

The results presented in the last section, and in particular the properties of $P_{eq}$, suggest that this classical model of the storage effect, with pure environmental noise, is incomplete and leads inevitably to a conceptual breakdown. Accordingly, an ``enrichment" of the storage dynamics with demographic noise is required.

As seen above, $P_{eq}$ depends only on $\delta$, the correlation time of environmental variations, and not on their strength $\sigma_E^2$ (as long as higher orders in $\sigma_E^2$ may be neglected). However, the environment is fluctuating on any timescale: the wind changes its velocity and direction, clouds cover the sky and cast a shadow, temperature changes slightly and so on. All these processes have correlation times of minutes or hours but their effect is minute. Still, under the rules that govern the lottery game, such processes have the ability to induce stability in the long run since $\delta \to 0$ and the amplitude of fitness fluctuations is irrelevant.

Clearly, a reasonable model should yield $P_{eq}$ that depends on $\sigma_E^2$. It is highly implausible that infinitesimal changes in wind direction or in temperature play an essential role in stabilizing natural communities, no matter how much time the system is allowed to relax. A lower cutoff below which environmental variations are negligible has to be introduced. For example, one may suggest that the minimal correlation time that the model has to take into account is the time between two consecutive deaths of individuals, since all the events that affect the fitness of species between two deaths are integrated to determine the success probability of every seedling competing to replace a dead tree. By doing that one already introduces the discreteness of individuals into the model.  Demographic stochasticity, which is the endogenous noise associated with the discreteness of the birth-death process, provides the natural mathematical tool to deal with these aspects of reality.  Quantifying the strength of demographic noise by the  parameter $\sigma_D^2$ (the value of this parameter is discussed below), we expect that the equilibrium distribution (\ref{eq3}) should be obtained, from a general theory with demographic fluctuations, in the limit $\sigma_D^2/\sigma_E^2 \to 0$, but for any finite demographic noise the ratio between  $\sigma_E^2$ and $\sigma_D^2$ should enter the expression for $P_{eq}$.

Another aspect of the result (\ref{eq3}), which also provides a hint about the importance of discreteness, is the transition between the single peak, balanced distribution at small values of $\delta$ and the imbalanced distribution at large $\delta$. Mathematically speaking, as long as the distribution is normalizable the solution is legitimate, so the theory holds for all $\delta <1$ and breaks down only when  $\delta=1$, where  $P_{eq}$ diverges like $x^{-1}$ at the edges.
However in practice, when the number of individuals has to be an integer, this formal approach may be misleading. If the overall size of the community is $J$ individuals, the case $x<1/J$ should be considered as extinction. No matter what $\delta$ is, in the long run one of the two species inevitably goes extinct and the system reaches fixation. This feature is missing in the lottery game, where all positive values are allowed for $x$.

Accordingly, the stability of the system depends not only on the shape of $P_{eq}$, but also on the rate at which the abundance of a single species  scans through all the values of $x$ and reaches values below $1/J$, a feature that depends strongly on $\sigma_E^2$  \cite{PhysRevE.92.022722}. Even if $P_{eq}(x)$ is very small for $x$ close to zero in the regime $\delta < 1/2$, under strong environmental noise the species' abundance samples the whole phase space on relatively short timescales, leading to fast fixation. As we shall see, since the decay of $P_{eq}$ at the edges is a power law at best, one cannot neglect extinctions even when $J$ is large.

Demographic stochasticity has two aspects. First, it opens the possibility of extinction by allowing a species to reach an absorbing state at zero concentration. Second, it provides another source of noise, which scales like the square root of the population size, as opposed to the linear scaling that characterizes environmental stochasticity \cite{kalyuzhny2014niche,lande2003stochastic}. These two aspects of demographic noise are of importance to the study of the storage effect, and they manifest themselves in TNTB.  However,  before considering TNTB  we would like to study the relevance of the storage effect, in its traditional form with only environmental stochasticity, to the statistics of high-diversity assemblages.

\section{The lottery game for many species }

The applicability of the storage effect as a possible explanation for an empirical system with tens and hundreds of species  was considered by Hubbell in \cite{hubbell_book}, during the introduction of the neutral model. The observed species abundance distribution in the tropical forest is very wide, with substantial support over a few decades of abundance; Hubbell argued  that the prediction for a system stabilized by the storage effect is a narrow SAD with a  Gaussian-like peak around some typical value. Accordingly, Hubbell concluded that the storage effect is inappropriate for explaining patterns of species diversity in the tropical forest. Since most of the diverse communities are characterized by a hollow curve of species abundances~\cite{ulrich2010meta} with many rare species and a few common ones, this argument suggests that the storage effect plays at best only a minor role in their dynamics.

In this section we will show that, in the limit of weak environmental stochasticity considered above,  when the number of species $S$ is much larger than one the storage effect yields a Gamma-like distribution for the SAD. The Gamma distribution is known to be mathematically flexible, it fits many empirical SADs and indeed it may resemble very closely the zero-sum multinomial distribution proposed by Hubbell.  Furthermore it contains the  commonly-observed Fisher log series SAD as a limiting case (see e.g., \cite{azaele2006dynamical, azaele2015towards,connolly2014commonness}).

\begin{figure}
\includegraphics[width=8cm]{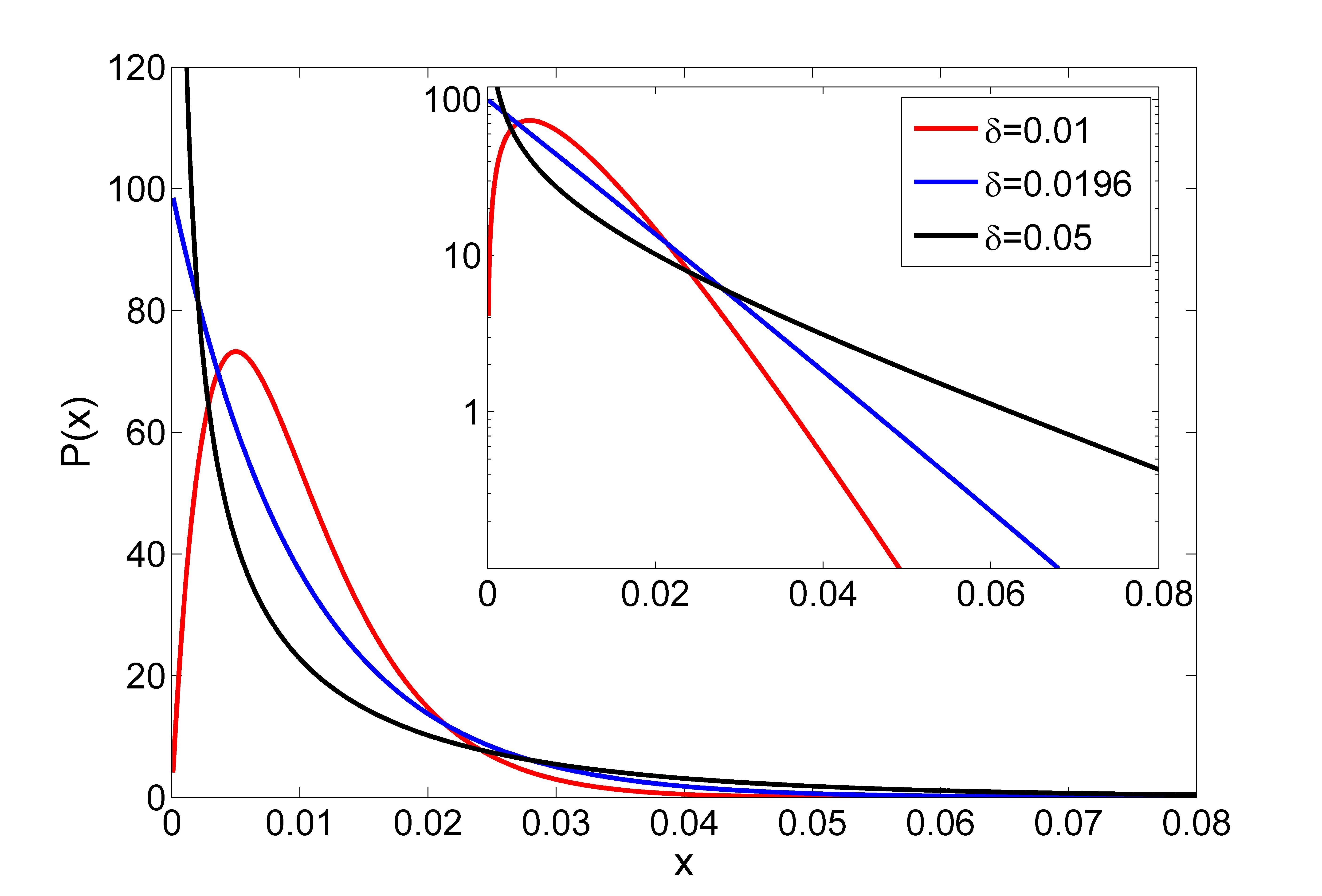}
\caption{The species abundance distribution $P_{eq}(x)$, given by Eq. (\ref{eq6}) for $S=100$ species, is plotted here vs. $x$ for three values of $\delta$. In all cases the SAD approximates a Gamma distribution, with shape factor $\alpha>1$ for $\delta<\delta_c=0.0196$ and $\alpha<1$ if $\delta > \delta_c \approx 0.0196$. At the critical $\delta$ the power vanishes and the distribution has almost pure exponential decay as demonstrated in the semi logarithmic plot in the inset. } \label{fig2}
\end{figure}

The generic, $S$ species,  generalization of Eqs. \ref{eq1} is,
\begin{equation}
x_i^{t+1} = (1-\delta) x_i^t + \delta \frac{\beta_i x^t_i}{\sum_{j=1}^S \beta_j x^t_j}.
\end{equation}
The solution for $P_{eq}$ was obtained by Chesson and Hatfield \cite{hatfield1997multispecies} (see also Gillespie \cite{gillespie1980stationary}),
\begin{equation} \label{eq5}
P_{eq}(x_1,x_2,...x_S) = (x_1 x_2 ...x_S)^{\frac{2}{S}(\frac{1}{\delta}-1)-1}.
\end{equation}
This is a very strong result, although it has not received its due attention in the literature.
To compare  directly the expression (\ref{eq5}) to observed SADs  one would like to extract the single species probability distribution function by integrating out $S-1$ species to obtain, say, $P_{eq}(x_1)$ (since all species are symmetric, we will denote it by $P_{eq}(x)$). The result is,
\begin{equation}\label{eq6}
P_{eq}(x) = x^{\frac{2}{S}(\frac{1}{\delta}-1)-1}
(1-x)^{\frac{2(S-1)}{S}(\frac{1}{\delta}-1)-1}.
\end{equation}

Eq. (\ref{eq6}) is an exact formula that reduces to (\ref{eq3}) where $S=2$, but in this section we are interested in its implication for $S \gg 1$. In this parameter regime $x \ll 1$, since the typical fraction of a single species never substantially exceeds $1/S$, as we shall see.

 As mentioned, when $S \gg 1$,  $P_{eq}$ takes the form of a Gamma distribution (power law followed by an exponential cutoff) form,
\begin{equation} \label{eq7a}
P_{eq} \approx x^{\alpha -1} e^{-\beta x},
\end{equation}
where the rate factor $\beta$ appears as the $x \ll 1$ limit of,
\begin{equation}
(1-x)^{\frac{2(S-1)}{S}(\frac{1}{\delta}-1)-1} \approx e^{-\beta x},
\end{equation}
 with  $\beta = (2/\delta)-3$. The shape factor $\alpha = \frac{2}{S}(\frac{1}{\delta}-1)$ will be greater than one (meaning that the distribution vanishes at zero and has a peak in the vicinity of $1/S$) if
\begin{equation} \label{eq8}
\delta < \delta_c = \frac{2}{S+2},
\end{equation}
 while for $\delta>\delta_c$ the distribution diverges at zero but is still integrable.  These behaviors are illustrated in Figure \ref{fig2}.  When $\alpha$ is small the distribution approaches the Fisher log series, and in general it corresponds to the generalized Fisher log series distribution that was discussed in \cite{kessler2014neutral}. Note that if $\delta>2/3$ the assumption $x \ll 1$ does not hold anymore; we will not consider this case here.


 For a fully surveyed empirical community the species richness $S$ is given, so the only parameter in (\ref{eq6}) which is left to be fitted is $\delta$. This makes the fit less impressive, of course, but the model is more parsimonious and its results may be preferred, e.g., when applying Akaike information criterion that includes a penalty to discourage overfitting, in comparison with two parameter theories.

\begin{figure*}
\includegraphics[scale=0.3]{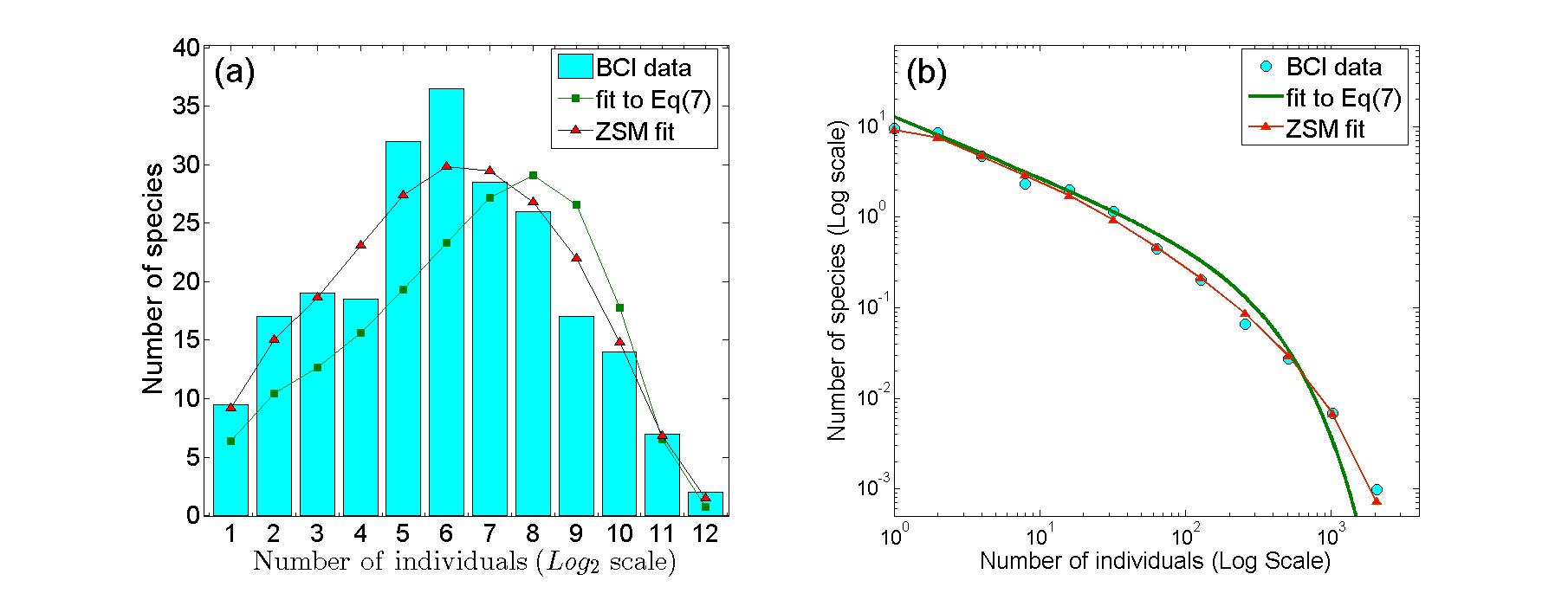}
\caption{Explaining the BCI plot species abundance distribution. The species abundance distribution of the BCI (1995 census) tropical forest is presented [light blue bars in the Preston plot (a), light blue circles in the double logarithmic (Pueyo) plot (b)], together with the best fit to the SAD predicted for a lottery game with $\delta = 0.0265$ (Eq. \ref{eq6}) and the ZSM theory (parameters taken from \cite{maritan1}, $\theta = 47.22$, $m=0.1$). Although the ZSM appears to give a better fit,  it is clear that Chesson-Warner lottery game does provide the kind of hollow curve SADs that appear in empirical studies.  Moreover, for the ZSM two parameters were used, whereas only one parameter was employed for the lottery game.} \label{fig4}
\end{figure*}

In particular, to explain the observed SAD on the BCI plot as reflecting a community that acquires its stability from the storage effect, we first look at Figure (\ref{fig4}b) and find the range of abundances for which the decay of the SAD is exponential, this indicates what $1/\beta$ is. In the BCI (dbh $ >10cm$) this "knee" (the range of abundances in which the SAD, on a double logarithmic scale as in \ref{fig4}b, curves down) is definitely below $500$ trees, i.e., below  $2\%$ of the forest population (around $21000$ trees). Since this scale is determined by $1/\beta$ where $\beta = (2/\delta)-3$, the  correlation time $\delta$ needed to explain the SAD in the Barro-Colorado plot has to be below $0.03$th of a generation. This estimate works quite nicely: Figure \ref{fig4} shows the observed SAD for BCI trees together with the $P_{eq}$ presented in equation (\ref{eq6}), where the only fitted parameter is $\delta$. Indeed the best fit was obtained for $\delta=0.027$, as expected. One can see that this one parameter fit is worse than the two-parameter fit using ZSM statistics, but it is not unacceptable. Interestingly, when plotted using the double-logarithmic (Pueyo) plot instead of Preston plot the single parameter fit using (\ref{eq6}) looks much better.

However, the value  $1/40$ (of a generation) for $\delta$ appears to be unrealistic. As mentioned in \cite{kalyuzhny2015neutral} the value of $\delta$ has been found to be around $1/5$, and the order of magnitude in difference is too large to be neglected.  Similarly, estimations of the corresponding  numbers for trees in the whole Amazon basin \cite{ter2013hyperdominance} suggest that the most abundant species constitutes about $1\%$ of the population (meaning that the ``knee" appears for even smaller relative abundances) and the corresponding value of  $\delta$, smaller than $ 1/200$, again seems unrealistic.

Altogether, it appears that a storage model may provide the type of hollow-curve SADs that characterize empirical systems. Its flexibility is limited since $\alpha$ and  $\beta$ are both determined by $\delta$,  and other theories provide better fits, still one may believe that adding another parameter (in a theory that takes into account spatial effects, for example) may solve this difficulty. The need to extend the lottery model and to include demographic noise is \emph{not} the inability of (\ref{eq6}) to support fat-tailed SADs, but the following three arguments:

 \begin{enumerate}
   \item Conceptually, as mentioned above, we would like to find an SAD that depends on $\sigma_E$, not only on $\delta$.
   \item The empirical SAD (and the theoretical expression (\ref{eq6}) with the $\delta$ values that yield a decent fit) has support on small absolute numbers, meaning that extinction events must occur and are important, or, equivalently, that demographic noise must be taken into account.
   \item The correlation time of the environmental variations needed to account for empirical datasets appears to be unrealistically short.
 \end{enumerate}

 In any case, once extinctions are incorporated into the model one should include speciation events to balance the species richness; the resulting model is the TNTB which is discussed in the next section.

\section{The TNTB: storage effect, demographic noise and speciation}

The time-averaged neutral theory of biodiversity deals with a community of species, all having the same \emph{average} fitness. Species are subject to demographic and environmental stochasticity. Under demographic noise species may go extinct, and these extinction events are balanced by speciation. Four parameters govern the results: in addition to  $\delta$ and $\sigma_E$, the correlation time and the amplitude of environmental variations, here one should take into account the per-birth chance of speciation  $\nu$, and the strength of the demographic noise, $\sigma_D$. One should introduce these two processes together: without extinction, speciation will cause the number of species to grow unboundedly. Without speciation, demographic noise will lead to fixation by a single species in the long run.

The standard way to introduce speciation is to assume that an offspring carries the taxonomic identity of its mother with probability $1-\nu$, and is the originator of a new species with probability $\nu$. The strength of demographic stochasticity is defined as the variance in the number of offspring per individual $\sigma_D^2$ and  usually takes a value between 2 (for a geometric distribution of offspring) and 1 (for a Poisson distribution). In the limit $\sigma_E = 0$, without environmental noise and storage effect, one obtains the metacommunity version of Hubbell's neutral theory (or Kimura's neutral model), where $P_{eq}$ (for a high-diversity system with $1/J < x \ll 1$ ) is given by Fisher log-series,
\begin{equation} \label{eq10}
P_{eq}^{\sigma_E = 0}(x) = \frac{A}{ x} e^{-\nu  J x/\sigma_D^2},
\end{equation}
where $A$ is a normalization constant. The  species richness $S$ reflects the balance between extinction  and speciation
\begin{equation} \label{eq11}
S^{\sigma_E = 0} = -\frac{\nu}{\sigma_D^2} \log\left(\frac{\nu}{\sigma_D^2}\right) J.
\end{equation}

What happens in TNTB, when the storage mechanism acts together with demographic noise and speciation events? Here we would like to emphasize a few generic features of this system:

\begin{enumerate}

  \item Demographic stochasticity allows for extinction while speciation increases species richness, and the balance between these two processes is determined by $\delta$ and $\sigma_E$. The lower the value of $\delta$, the sharper is the SAD peak in the vicinity of $1/S$, the time to extinction of a single species increases, and the chance of a low-abundance species to invade grows. Accordingly, the steady state species richness $S$ decreases monotonically with increasing $\delta$.

\begin{figure}
\includegraphics[width=8cm]{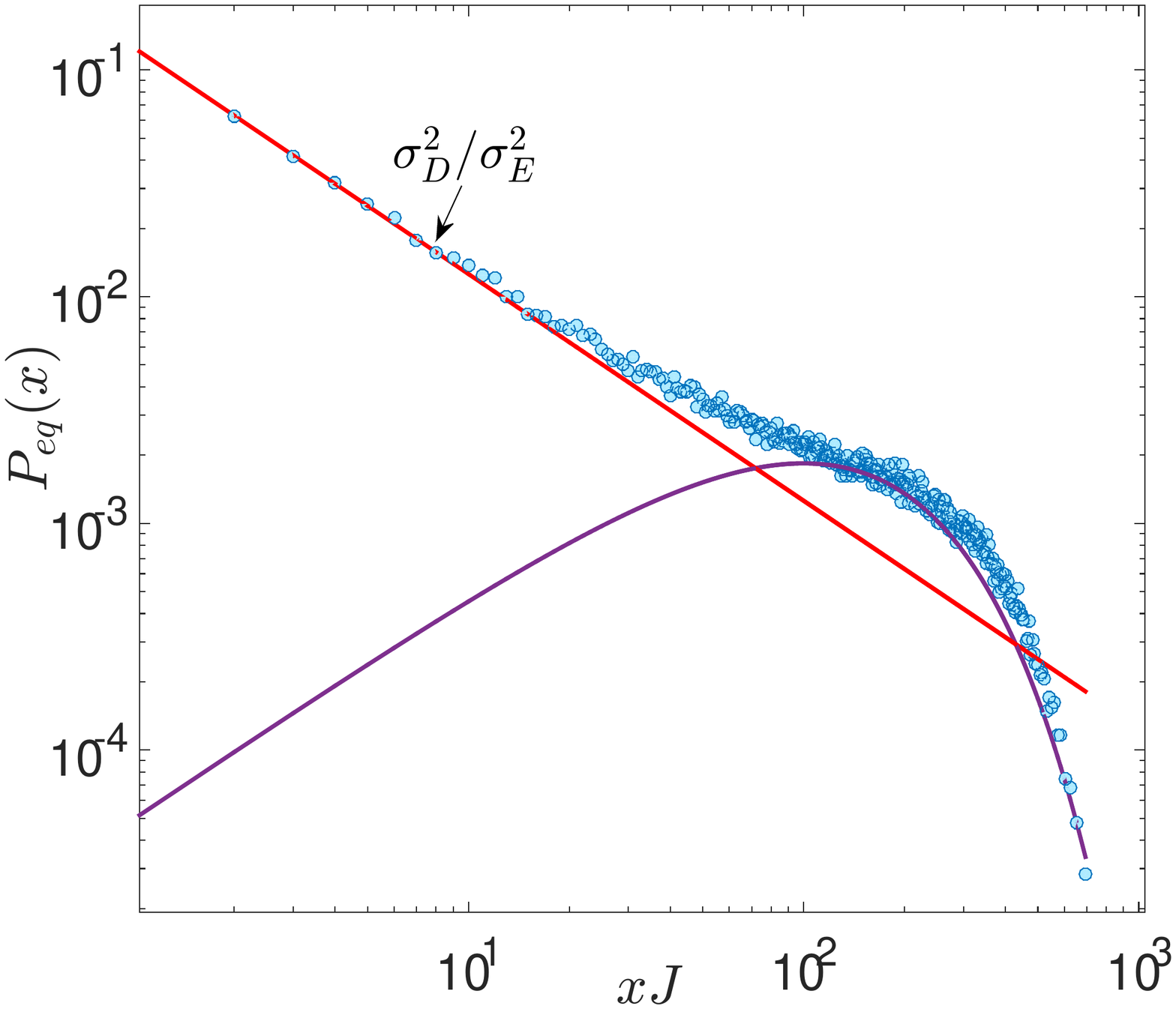}
\vspace{-5cm}
\caption{The light blue circles represent the  SAD obtained from a simulation of the TNTB model, with $\sigma_D^2=2,\ \sigma_E^2=0.25,\ \delta=10^{-4},\ \mu=10^{-3}\ and\ J=10^4$. The results represent an average over time, abundances have been recorded every $10^6$ elementary timesteps. The red straight line has a slope $-1$, and it fits the data perfectly up to $\sigma_D^2/\sigma_E^2$, the region dominated by demographic stochasticity. Gamma distribution, with $\alpha = 2$ and $\beta = 0.01$ is shown by the purple curve.} \label{fig5}
\end{figure}

      Dynamically, when the chance of extinction is low the process of speciation acts to increase the number of species $S$, $\delta_c$ decreases (see Eq. \ref{eq8}) and the support of $P_{eq}$ in the region $x \ll 1$ grows, leading to an increased rate of extinction until it balances the effect of speciation  and the system reaches a steady state at finite $S$.

  \item The inclusion of both demographic and environmental noise introduces a new scale into the problem. As discussed in \cite{kessler2014neutral,PhysRevE.92.022722}, as long as $x < \sigma_D^2/\sigma_E^2$ the dynamics of a species is dominated by demographic noise, while, above this value, environmental variations are more important. Accordingly, as one can see in Figure \ref{fig5}, $P_{eq}$ of the TNTB has two regimes.  For large $x$ one observes  (\ref{eq7a}), the storage power-law $\alpha-1$ followed by an exponential cutoff.  For $x < \sigma_D^2/\sigma_E^2$ the power $\alpha-1$ is replaced by a $1/x$ dependence, a characteristic of the Fisher log-series.

Therefore, the conceptual problem raised in Section III, namely the fact that the theory of the storage effect predicts $P_{eq}$ to be independent of $\sigma_E$, is solved within the TNTB framework: the ratio $\sigma_D^2/\sigma_E^2$ determines the crossover from the $1/x$ decay to the behavior described by (\ref{eq7a}), and the SAD (and the overall species richness $S$) does depend on $\sigma_E$. In the $\sigma_E \to 0$ limit the TNTB converges to the standard neutral theory of Hubbell.

\begin{figure}
\vspace{-2cm}
\includegraphics[width=8cm]{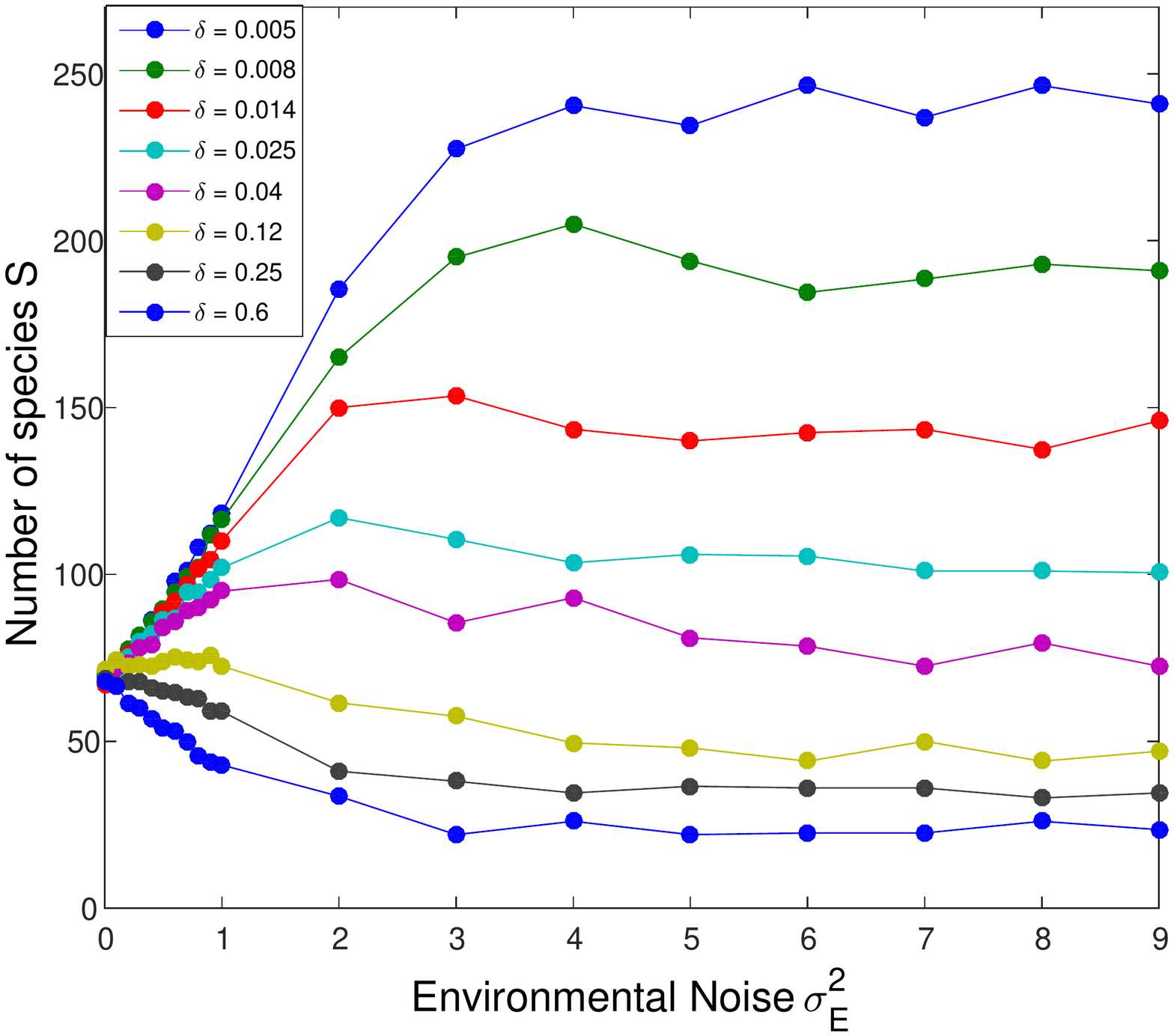}
\vspace{-2cm}
\caption{$S$, the species richness, is plotted against $\sigma_E^2$, the amplitude of the environmental stochasticity. The results were obtained in simulations of a TNTB community with $J = 10000$ and $\nu= 0.001$ for various environmental correlation times  $\delta$ (given in the legend in units of one generation). $S$ reflects the balance between extinction and speciation; the lower is $\delta$, the stronger is the storage effect and thus $S$ increases. An increase in the strength of environmental variation $\sigma_E^2$ may either decrease $S$ (since it increases abundance variation)  or increase the species richness by facilitating the storage effect. Here we see that the general trend  depends on the value of $\delta$.  All the lines converge to the NTB limit, $S = -J\nu \log(\nu) \cong 70$, when $\sigma_E \to 0$.} \label{fig6}
\end{figure}

  \item Given that, one may wonder about the effect of  environmental stochasticity on species richness. On the one hand, $\sigma_E$ is responsible for the storage effect that provides stability and allows for low abundance species to invade. On the other hand, (see Eq. (\ref{eq12}) below and the following discussion) in systems without a storage effect \cite{kessler2014neutral,PhysRevE.92.022722}, environmental stochasticity clearly acts to lower $S$, as it increases the rate of extinction events  since environmental  fluctuations  cause a species to visit more frequently the dangerous zone of low abundance.

      Figure \ref{fig6} solves this puzzle: it shows that the effect of environmental stochasticity on species richness, when all other parameters are kept fixed, is determined by the correlation time $\delta$. For small $\delta$'s the storage effect wins and in general the species richness increases with the amplitude of environmental variations. For large values of $\delta$ the increase in the system's variability leads to a decrease in $S$.

      \item The NTB was criticized by many authors for its strict commitment to perfect neutrality \cite{zhou2008nearly}. Under the rules of the neutral game, even the slightest fitness difference leads to a fixation of the system by the fittest species (in the absence of speciation) or to the appearance of an SAD that reflect Darwinian dominance, with one common species that occupies most of the community and a few rare, short lived, species \cite{kessler2014neutral}. The stabilizing effect of the storage mechanism resolves this difficulty.  Even if the  average fitness of different species is \emph{not} the same, the system may still support high diversity.

          To demonstrate this we have simulated  the non-neutral modification of the TNTB, when the expression for  fitness,  Eq. (\ref{eq0}), is replaced by
           \begin{equation} \label{eq01}
      f_i = e^{\eta_i + \gamma_i^t},
       \end{equation}
       where $\eta_i$ is a time independent, species specific component of the fitness of the $i$-th species, taken from a Gaussian distribution with standard deviation $\mu$ and zero mean. $\eta_i$ reflects the mean tendency of the environment to favor, or disfavor, species $i$. When $\sigma_E \to 0$ the introduction of these time independent fitness differences leads to a biodiversity collapse, as seen in Figure \ref{fig7}. However, as $\sigma_E$ increases, the number of species grows since the storage effect induces stability.  Finally at large  $\sigma_E$  the effect of fitness differences disappears and the species richness takes its $\mu=0$ value, as in the TNTB.

       An interesting feature of the finite $\mu$ dynamics is the unimodal dependence of $S$ on $\sigma_E^2$ when $\delta$ is sufficiently large. While weak environmental stochasticity stabilizes the species, strong variations lead to faster extinction and reduce $S$ and biodiversity reaches a maximum under \emph{intermediate disturbance}. One may expect such an effect in nonadditive systems (see \cite{fox2013intermediate}), and we believe that our model provides an appropriate framework for its analysis.

\begin{figure}
\vspace{-2cm}
\includegraphics[width=8cm]{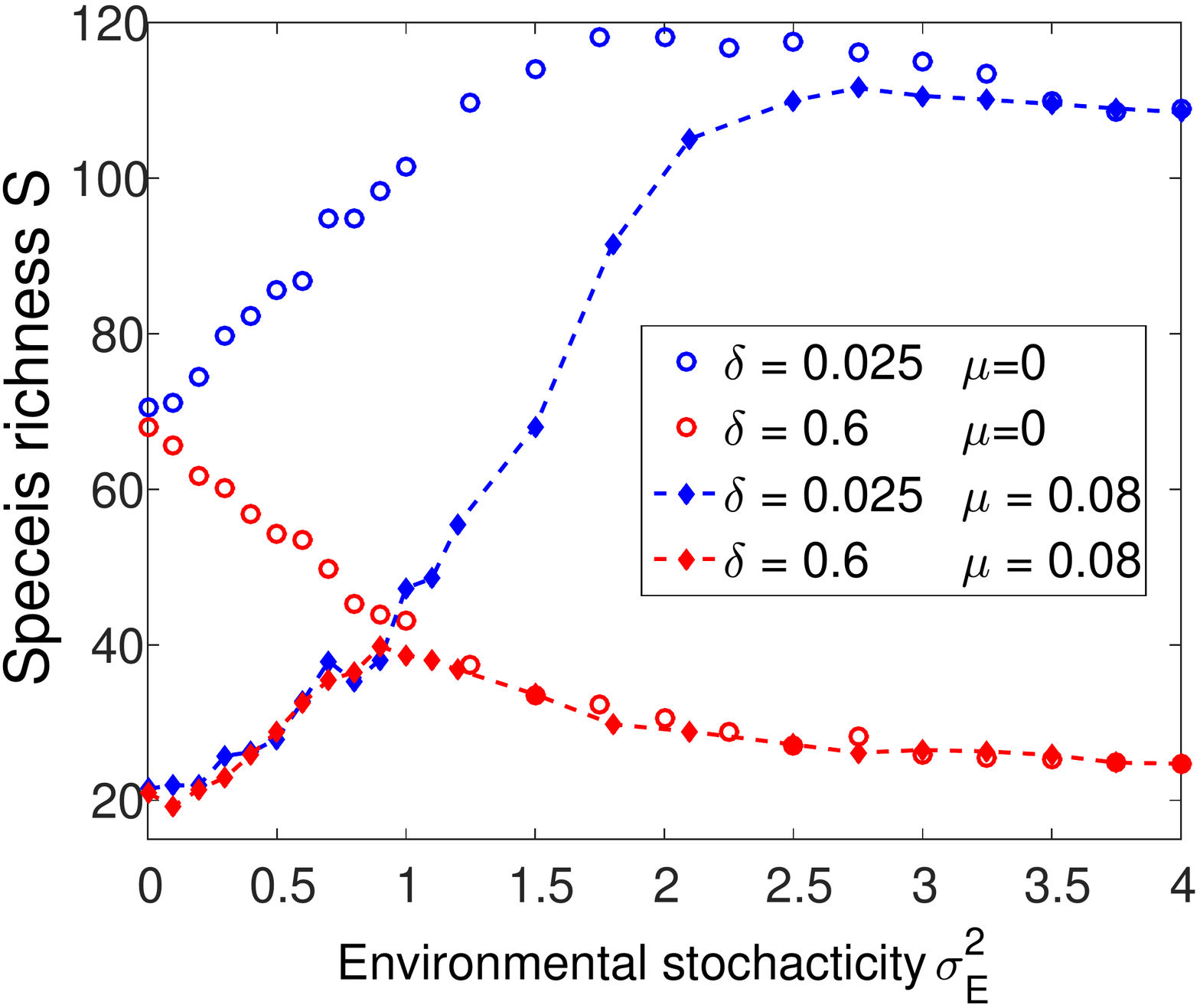}
\vspace{-2.2cm}
\caption{$S$, the species richness, is plotted against $\sigma_E^2$, amplitude of the environmental stochasticity for simulations with $\mu=0$ (same as in Fig \ref{fig6}, empty circles) and for the case with time independent fitness differences ($\eta_i$s) with $\mu=0.08$ (filled circles connected by dashed lines). All other parameters are the same used in Figure \ref{fig6}.   When $\sigma_E^2 =0$ time independent fitness differences lead to a biodiversity collapse, where the fittest species is dominant and all other species are rare. On the other hand, when $\sigma_E^2$ is large environmental noise washes out the effect of time independent fitness differences and $S(\mu)$ approaches $S(\mu=0)$. The species richness peaks at intermediate level of disturbance for $\delta = 0.6$.} \label{fig7}
\end{figure}

\end{enumerate}

 Before concluding this section, we would like to stress that environmental stochasticity and the storage effect are not synonymous. When the environmental variations affect only  the death rate, or when $\delta = 1$,  there is no storage effect and $\sigma_E$ is a purely destabilizing factor. Even a slight modification of the rules governing the process may kill the storage effect. For example, in the lottery game \emph{all} the trees in the forest are competing for an open gap, where the chance of a species to win depends on its fitness. The pairwise competition version of the same game, where two individuals are picked at random and an offspring of one replaces the other with probability that depends on their relative fitness, has no significant storage effect. With respect to this specific duel all other trees play no role, so it corresponds to the $\delta=1$ case of the lottery game. Conversely, in the NTB limit (without environmental stochasticity) there is no difference between these two versions of the neutral game, see e.g. \cite{maritan1}.

The SAD for TNTB \emph{without} the storage effect (e.g., for the pairwise competition case) for $1/J<x \ll 1$ was calculated in \cite{kessler2014neutral},
\begin{equation} \label{eq12}
P_{eq}^{{\rm \ no \  storage}}(x) = \frac{C}{x}\left(1+\frac{2\sigma_E^2 J }{\sigma_D^2}x\right)^{-1-\nu/\sigma_E^2}
\end{equation}
where  $C$ is a normalization constant. Here  the  power law decay $1/x$ that characterizes the region dominated by demographic noise  is replaced, for $x>\sigma_D^2/\sigma_E^2$, by a power law with a larger exponent $2+\nu/\sigma_E^2$.

\section{Summary and discussion}

Mechanisms that maintain species diversity are usually classified according to their stability properties. Some mechanisms provide a stable equilibrium, while in other mechanisms  the dynamics of each species is unstable and the diversity reflects a balance between extinction and speciation/immigration. This distinction is related to timescales: under the inevitable influence of demographic noise every species eventually goes extinct, however in models that support a stable equilibrium the extinction time is exponential in the species' abundance, while under unstable equilibrium, like in the neutral model, the time to extinction scales linearly with the abundance. To maintain the diversity of a metacommunity these timescales should be comparable with the evolutionary timescale that determines the rate at which new species enter the system and balance the diversity losses due to extinction.

A system that acquires its stability due to the storage effect is somewhere in-between. The stabilization is based on environmental stochasticity, which is, at the same time, a destabilizing force.  As we have seen, the outcome of the competition between these two aspects of the same phenomenon - environmental stochasticity - is determined by one parameter, $\delta$, the correlation time of the environment. If $\delta$ is large the destabilizing effects dominate and environmental stochasticity reduces biodiversity. When $\delta$ is small, as seen in fig \ref{fig6}, the stabilizing effect associated with the storage mechanism leads to an increase of extinction times and the overall biodiversity.

In the metacommunity version of Hubbell's neutral theory, speciation and demographic drift are the only factors that govern the dynamics of the community, leading to the Fisher log-series SAD and species richness which is given by (\ref{eq11}).  The $1/x$ decrease of the SAD at small $x$ does not fit the observed statistics on, say, the Barro Colorado Island and other local communities, where the slope is clearly weaker than $1/x$ (in a Preston plot, where the number of species in any abundance octave is plotted without normalization by the width of the octave, $1/x$ is translated into a straight horizontal line, while the Preston plots of empirical local communities show a unimodal behavior, see Fig \ref{fig4}a). To account for that, in the mainland-island version of NTB the statistics of a local community are governed by two parameters, the fundamental biodiversity number of the metacommunity $\theta = 2 \nu J_M /\sigma_D^2$  and the chance of migration to the mainland. The emerging zero-sum multinomial  SAD fits the empirical evidence, as may be seen in Figure \ref{fig4} above.  Nevertheless  the dynamics, in particular the rate of abundance variations, is too fast to be explained by the neutral model~\cite{kalyuzhny2014niche,  kalyuzhny2014temporal,  kalyuzhny2015neutral}.

TNTB, that was shown to explain both static and dynamic patterns~\cite{kalyuzhny2015neutral}, has three extreme limits. When $\sigma_E \to 0$ it converges to the NTB, as environmental variations vanishes. When $\sigma_D/\sigma_E \to 0$ it converges to the classical lottery model of Chesson and Warner. The other limit is $\delta \to 1$, when environmental noise does affect the system but there is no storage. The SADs in these three limits were presented in this paper (Eqs. \ref{eq6}, \ref{eq10}, \ref{eq12}).

In between, as showed in section IV, the situation is more complicated, and the way environmental stochasticity affects  species richness is determined by the correlation time $\delta$. For short correlation times, $S$ is an increasing function of $\sigma_E$, while for longer correlation times the situation is closer to the one discussed in \cite{kessler2014neutral} - a species may enjoy a long time in which its population grows, so the SAD widens and the overall species richness $S$ decreases when environmental variations increase in amplitude.

Finally, we would like to comment about the concept of ``speciation" as used through this paper. In the original neutral theory  all individuals are identical, so speciation has no relevant meaning other than a statement about the phylogenetic heritage of an individual.  The theory puts no constraints on taxonomic classification, which may be based on any property, from genetic sequences to eye color to beak size. The lottery game, and the storage effect associated with it, requires differential response of a species to the varying conditions, meaning that two species that respond in the same way to temperature or precipitation should be considered as a single species in the lottery game, despite having pronounced phenotypic differences. Accordingly, the collection of species in, say, a tropical forest, may admit two levels of taxonomic classification from the viewpoint of the TNTB:  species that respond together to the environment are playing an NTB  game among themselves, and the lottery game is played, not between species but among different functional groups. Accordingly, ``speciation" in the TNTB model is not necessarily equivalent to the birth of a new species in the sense of traditional systematics,  it is related to different response to the environment. This feature may shed a new light on the evolutionary process, as reflecting the long-term outcomes of community dynamics.

\bibliography{chesson_ref}

\end{document}